%% file: 19C13_main.tex
\documentclass[preprint]{raa}            
\usepackage{natbib}
\usepackage{braket}
\usepackage{graphicx,times}             
\usepackage{amssymb,amsmath}
\usepackage[pagebackref=true]{hyperref}
\usepackage{ulem}
\usepackage{threeparttable}
\usepackage{longtable}

\bibpunct{(}{)}{;}{a}{}{,}

\newcommand{\Bmean}{$\langle B_{\parallel}\rangle$}

\usepackage{color}

\begin{document}

\title{Reciprocating Magnetic Fields in the Pulsar Wind Observed from the Black Widow Pulsar J1720$-$0534}

\volnopage{Vol.0 (20xx) No.0, 000--000}      
\setcounter{page}{1}          

\author{Chen-Chen~Miao 
  \inst{1,2}
\and Victoria~Blackmon 
  \inst{3}
\and Wei-Wei~Zhu 
  \inst{1,4}
\and Dong-Zi~Li 
  \inst{5}
\and Mingyu~Ge %
  \inst{6}
\and Xiao-Peng~You 
  \inst{7}
\and Maura~McLaughlin 
  \inst{3}
\and Di~Li 
  \inst{1,8}
\and Na~Wang 
  \inst{9}
\and Pei~Wang 
  \inst{1,4}
\and Jia-Rui~Niu  
  \inst{1,2}
\and M.~Cruces
  \inst{10}
\and Jian-Ping~Yuan 
  \inst{9}
\and Jun-Tao~Bai 
  \inst{9}
\and D.~J.~Champion
  \inst{10}
\and Yu-Tong~Chen 
  \inst{1,2}
\and Ming-Min~Chi 
  \inst{11}
\and P.~C.~C.~Freire
  \inst{10}
\and Yi~Feng 
  \inst{8}
\and Zhen-Ye~Gan 
  \inst{15}
\and M.~Kramer
  \inst{10}
\and Fei-Fei~Kou 
  \inst{9}
\and Yu-Xi~Li 
  \inst{15}
\and Xue-Li~Miao 
  \inst{1}
\and Ling-Qi~Meng 
  \inst{1,2}
\and Chen-Hui~Niu 
  \inst{1,13}
\and Sheng-Nan~Sun 
  \inst{9}
\and Zhong-Yi~Sun 
  \inst{16}
\and H.~M.~Tedila 
  \inst{9}
\and Shuang-Qiang~Wang 
  \inst{9}
\and Qing-Dong~Wu 
  \inst{9}
\and Jing-Bo~Wang 
  \inst{14}
\and Zhi-Gang~Wen 
  \inst{9}
\and Shen~Wang 
  \inst{11}
\and Ya-Biao~Wang 
  \inst{15}
\and Cheng-Jie~Wang 
  \inst{15}
\and Meng-Yao~Xue 
  \inst{1}
\and You-Ling~Yue 
  \inst{1}
\and Mao~Yuan 
  \inst{1,2}
\and Ju-Mei~Yao 
  \inst{9}
\and Wen-Ming~Yan 
  \inst{9}
\and Ru-Shuang~Zhao 
  \inst{1,12}
\and Lei~Zhang 
  \inst{1}
\and De~Zhao 
  \inst{9}
}

\institute{National Astronomical Observatories, Chinese Academy of Sciences, Beijing 100012, China; {\it zhuww@nao.cas.cn}\\ 
\and 
School of Astronomy and Space Science, University of Chinese Academy of Sciences, Beijing 100049, China\\
\and 
West Virginia University, 1500 University Ave, Morgantown, WV 26506, USA \\
\and 
Institute for Frontiers in Astronomy and Astrophysics, Beijing Normal University, Beijing 102206, China\\
\and 
Cahill Center for Astronomy and Astrophysics, California Institute of Technology, 1216 E California Boulevard, Pasadena, CA 91125, USA \\
\and 
Key Laboratory of Particle Astrophysics, Institute of High Energy Physics, Chinese Academy of Sciences, Beijing 100049, China\\
\and 
School of Physical Science and Technology, Southwest University, Chongqing 400715, China\\
\and 
Research Center for Intelligent Computing Platforms, Zhejiang Laboratory, Hangzhou 311100, China\\
\and 
Xinjiang Astronomical Observatory, Chinese Academy of Sciences, Urumqi, Xinjiang 830011, China\\
\and 
Max-Planck Institut f{\"u}r Radioastronomie, Auf dem H{\"u}gel 69, D-53121 Bonn, Germany\\ 
\and 
School of Computer Science, Fudan University, Shanghai 200433, China\\
\and 
Guizhou Normal University, Guiyang 550001, China\\
\and 
College of Physical Science and Technology, Central China Normal University, Wuhan 430079, China\\
\and 
Institute of Optoelectronic Technology, Lishui University, Lishui, Zhejiang,323000, China\\
\and 
Tencent Youtu Lab, Tencent Youtu Lab, Shanghai, 200233, China\\\vs\no
   {\small Received 20xx month day; accepted 20xx month day}}

\abstract{
We report the radio observations of the eclipsing black widow pulsar J1720$-$0534, a 3.26~ms pulsar in orbit with a low mass companion of mass 0.029 to 0.034~$M_{\odot}$.
We obtain the phase-connected timing ephemeris and polarization profile of this millisecond pulsar (MSP) using the Five-hundred-meter Aperture Spherical Radio Telescope (FAST), the Green Bank Telescope (GBT), and the Parkes Telescope.
For the first time from such a system, an oscillatory polarisation angle change was observed from a particular eclipse egress with partial depolarization, indicating 10-milliGauss-level reciprocating magnetic fields oscillating in a length scale of 5$\times10^3$~km (assuming an orbital inclination angle of 90 degrees) outside the companion's magnetosphere.
The dispersion measure variation observed during the ingresses and egresses shows the rapid raising of the electron density in the shock boundary between the companion's magnetosphere and the surrounding pulsar wind. 
We suggest that 
the observed oscillatory magnetic fields originate from the pulsar wind outside the companion's magnetosphere.
}

\authorrunning{C.-C.~Miao}            
\titlerunning{BW pulsar Wind Magnetic Fields}  
\maketitle


\section{Introduction}
Spider pulsars are millisecond pulsars (MSPs) orbiting low-mass companions ($M_{\rm C}<~1~M_{\odot}$) in tight orbital periods ($P_{b} < 24$~hr) \citep{Roberts+2011, Roberts+2013IAUS}.
There are two explicit subgroups of spider pulsars \citep{Roberts+2011}: black widows (BWs) have likely-degenerate companions with mass $M_{\rm C}<0.05~M_{\odot}$, and redbacks (RBs) have non-degenerate companions with mass $M_{\rm C}\sim0.1-1.0~M_{\odot}$. 
Spider pulsar systems commonly show eclipsing behaviours like: frequency-dependent radio intensity eclipse wider than the companion physical size \citep{Fruchter+1988, Polzin+2018MNRAS} and linear polarization eclipse leading and trailing the intensity eclipse by a few degrees \citep{you+2018ApJ, li+2022arXiv220507917L, Crowter+2020MNRAS}.

The magnetic field in the eclipse medium causes the polarization eclipse of spider pulsars \citep{Thompson+1994ApJ, Kansabanik+2021ApJ}.
As suggested by \cite{Thompson+1994ApJ} and \cite{Polzin+2018MNRAS}, the eclipsing materials should contain $\gtrsim$10~G magnetic field to balance the energy density in the pulsar relativistic-particle wind. 
The recent result of PSR~J1544+4937 also shows that eclipsing material needs to contain $\sim$10~G magnetic field
from wide-band flux intensity modeling \citep{Kansabanik+2021ApJ}.
However, observations of the depolarization eclipse do not always agree with the prediction above.
Based on the plasma lensing argument, the average magnetic field parallel to the line-of-sight (LOS) {\Bmean} for PSR~B1957+20 is $0.02\pm0.09~$G \citep{li+2019MNRAS}.
\cite{Crowter+2020MNRAS} measure a corresponding $B_{\parallel}\sim3.5\pm1.7~$mG at the eclipse egress of PSR~J2256$-$1024 through polarization angle changes caused by Faraday rotation.
Other indirect evidences point to a higher magnetic field strength in the eclipsing materials.
\cite{li+2022arXiv220507917L} measure substantial RM changes over particular orbital phases away from the eclipse and changes of circular polarization consistent with the synchrotron-cyclotron absorption caused by magnetic fields of 10-100~G from PSR~1744-24A.

PSR~J1720$-$0534 \citep{wang+2021ApJ} is a Galactic field BW pulsar with a spin period of 3.26~ms discovered in the Commensal Radio Astronomy FAST Survey (CRAFTS) \citep{lwq+18, Cameron+2020MNRAS, Cruces+2021MNRAS, Miao+2023MNRAS}.
The discovery of this BW pulsar and the presentation of pulse intensity variation with a modulation period of $\sim$22~s during the ingress of the eclipse has been reported \citep{wang+2021ApJ}.
In this work, we report the phase-connected timing ephemeris, polarization profile, and a magnetic field reversal of this pulsar. 
In Section \ref{sec:process}, we present the radio observations and data analyses.
In Section \ref{sec:result}, we describes our timing results, linear polarization variation with the orbital phase, and a only magnetic field reversal in an observation.
We discuss and summarize our results in section \ref{sec:discuss}.

\section{Observation and Data Analysis}\label{sec:process}
\subsection{Observation}
There were 42 observations performed at FAST with its 19-beam L-band receiver \citep{Nan+2011IJMPD, jth+20}, which covers a frequency band of 1.0$-$1.5~GHz.
To ensure the possible smallest data sizes, we used two different observation modes.
The regular timing observations were carried out with 1024 frequency channels and 49.152~${\mu}s$ sampling time. 
The analysis of DM and RM variation was carried out with 8192 frequency channels and a sampling time of 98.304~${\mu}s$.
The observations at the 100-m Green Bank Telescope (GBT) were carried out using the 350~MHz and 820~MHz feeds at its prime focus receiver. 
Data were coherently de-dispersed using the VEGAS backend at a DM of 37.8~pc~cm$^{-3}$ with 128 frequency channels, and sampling times of 20.48~${\mu}$s and 10.24~${\mu}$s at 350 and 820 MHz, respectively.
The observation taken at the Parkes telescope was made with the ultra-wide-bandwidth low-frequency (UWL) receiver \citep{parkesUWB+20}. This 2.7~hours observation covered $\sim$90~percent of the orbital phase.
All the data presented in this work were recorded in pulsar search mode with full polarimetry.
The parameters of timing observations and eclipsing events are listed in Tab.~\ref{tab:obsLog}.

\subsection{Data Processing}
The initial timing ephemeris was derived with tools from \texttt{PRESTO}\footnote{\url{https://github.com/scottransom/presto}\label{fn:presto}} \citep{ran11}.
We folded the data with the initial ephemeris by using \texttt{DSPSR}\footnote{\url{http://dspsr.sourceforge.net}} \citep{vb11} and obtained the pulse time of arrivals (ToAs) with \texttt{PSRCHIVE}\footnote{\url{http://psrchive.sourceforge.net}\label{fn:psrchive}} \citep{hvm04} software packages.
We used \texttt{TEMPO}\footnote{\url{http://tempo.sourceforge.net}} \citep{Nice+2015_tempo} and \texttt{DRACULA}\footnote{\url{https://github.com/pfreire163/Dracula}} \citep{fr18} to get a phase-connected timing solution with the initial ephemeris and these ToAs.
After obtaining a phase-connected timing solution, we refolded the data and performed the polarimetric calibration on the newly folded data.
We derived the new ToAs with \texttt{PSRCHIVE} and get the timing residuals by using \texttt{TEMPO}.
ToA integration times used for FAST, GBT, and Parkes observations are is the 20~s, 120~s, and 420~s, respectively.

\subsection{Faraday Rotation Measurements}
The observations covering the eclipses are all taken at 1250~MHz with the FAST.
To study the depolarization of the pulsar signals during the eclipse phase, we search for the optimal $\rm{RM}$ using \texttt{RM-TOOLS}\footnote{\url{https://github.com/CIRADA-Tools/RM-Tools}} \citep{pvw+20} from each sub-integration.
We obtained polarization profiles with the signal-to-noise ratio $\sim$ 80 by forming profiles for every 20~s.
We find the $\rm{RM}\approx21~\rm{rad~m^{-2}}$ at the non-eclipsing phases ($0.00<\phi_{b}<0.15$ and $0.40<\phi_{b}<1.00$).

The \texttt{RM-TOOLS} failed to provide a fit result for the ingress and egress phase ($0.15<\phi<0.18$ and $0.31<\phi_{b}<0.40$), due to the low linear polarization.
We then attempt to obtain the RM variation by measuring the polarization position angle (PA) shift of the sub-integration \citep{Crowter+2020MNRAS}.
The extra RM is given by $\Delta \rm{RM}=\Delta \rm{PA}$${f_{c}^{2}c^{-2}}$, where $\Delta \rm{PA}$ is the PA shift; $c$ is the speed of light in $\rm{m~s^{-1}}$; and $f_{c}=1250$~MHz is the center frequency of the observation.
We compare PA values between the shifted and non-shifted PA profiles to obtain the $\Delta \rm{PA}$ measurements.
\section{Results}\label{sec:result}

We present a phase-connected timing solution for PSR~J1720$-$0534 (Tab.~\ref{tab:par}) based on two years of observations.
This 3.26~ms pulsar is in a tight orbit with an orbital period of 3.16 hours, an eccentricity of $4.4\times10^{-5}$, and a projected semi-major axis of 0.0596~lt-s.
The companion mass is to be between 0.029$-$0.068~$M_{\odot}$, assuming $90^{\circ}\geq i\geq 26^{\circ}$ and a pulsar mass of 1.35~$M_{\odot}$.
We derive the spin-down luminosity (${\dot{E}}$), the surface magnetic field (${B_\text{surf}}$), and the characteristic age (${\tau_\text{c}}$) from the intrinsic period (${P}$) and spin period derivative (${\dot{P}}$).
The ${\dot{E}}$ of $9\times10^{33}$~erg~s$^{-1}$ is similar to those of the 102 published LAT-detections of Gamma-Ray MSPs\footnote{\url{https://confluence.slac.stanford.edu/display/GLAMCOG/Public+List+of+LAT-Detected+Gamma-Ray+Pulsars}} with measured ${\dot{E}}$.
We folded the 12-year Fermi-LAT data with the timing solution in Tab.~\ref{tab:par}, but no pulse was detected.
We searched the potential counterpart at the Gaia DR3 catalogues \citep{Gaia+2021A&A} and none was found within a 1 arcsec radius.

\begin{figure}
\centering
\includegraphics[width=0.95 \columnwidth]{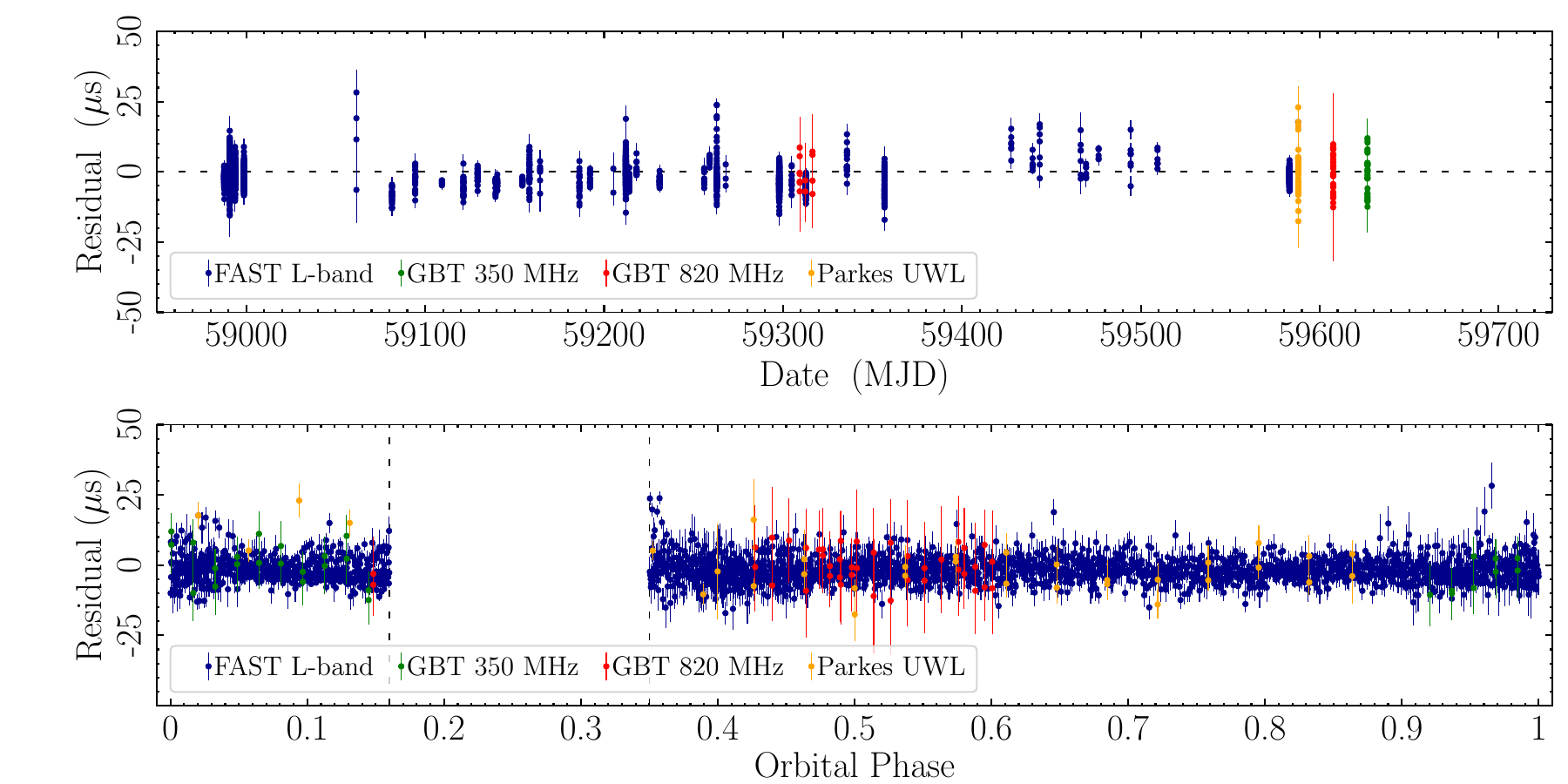}
\caption{
Post-fit timing residuals of all observations.
Lower panel: The timing residuals vs. MJD.
Bottom panel: The timing residuals vs. orbital phase.
Residuals at 350~MHz and 820~MHz from GBT observations are plotted in green and red, while residuals from FAST and Parkes observations are plotted in blue and yellow, respectively.
The vertical lines in the lower panel represent the approximate location of the eclipse, at orbital phases between 0.16 and 0.35.
}
\label{fig:res}
\end{figure}

\begin{figure}
\centering
\includegraphics[width=0.95 \columnwidth]{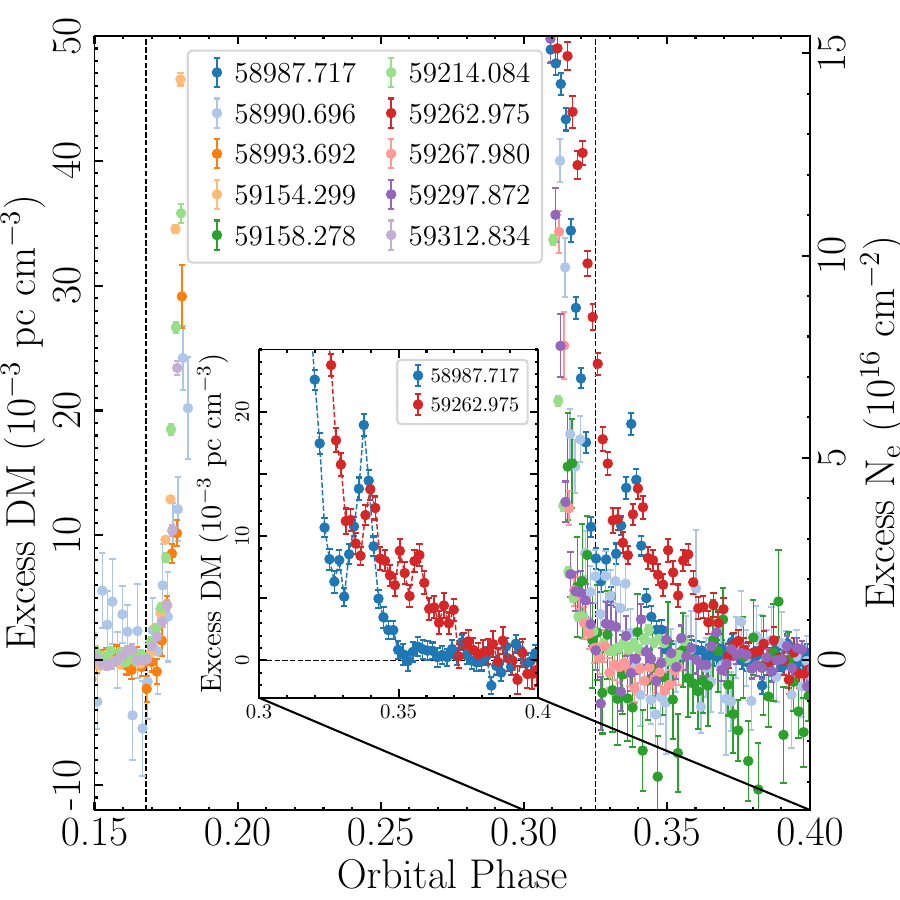}
\caption{Variations of the excess DM (${\Delta}$DM) and excess electron column densities ($N_{e}$) as the function of orbital phases around  superior conjunction. Different colors represent the different eclipse observations. The inset shows the DM "blips" observed on MJD~58987 and MJD~59262.
}
\label{fig:dmx}
\end{figure}

\begin{figure}
\centering
\includegraphics[width=0.49 \columnwidth]{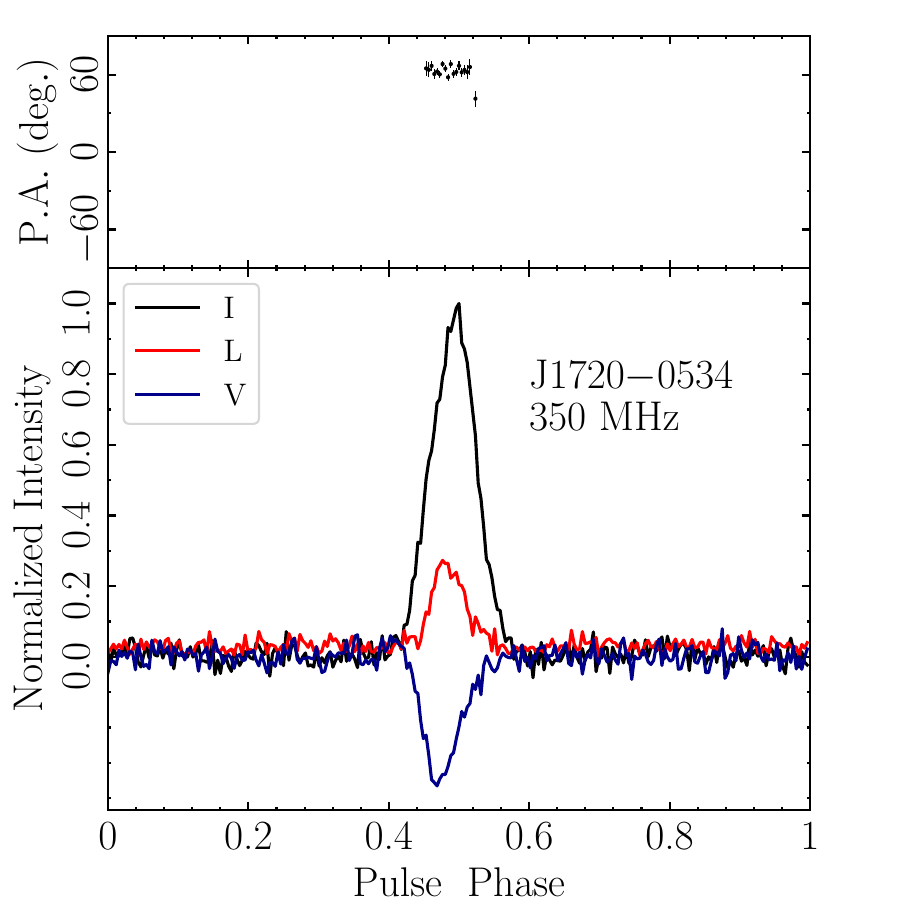}
\includegraphics[width=0.49 \columnwidth]{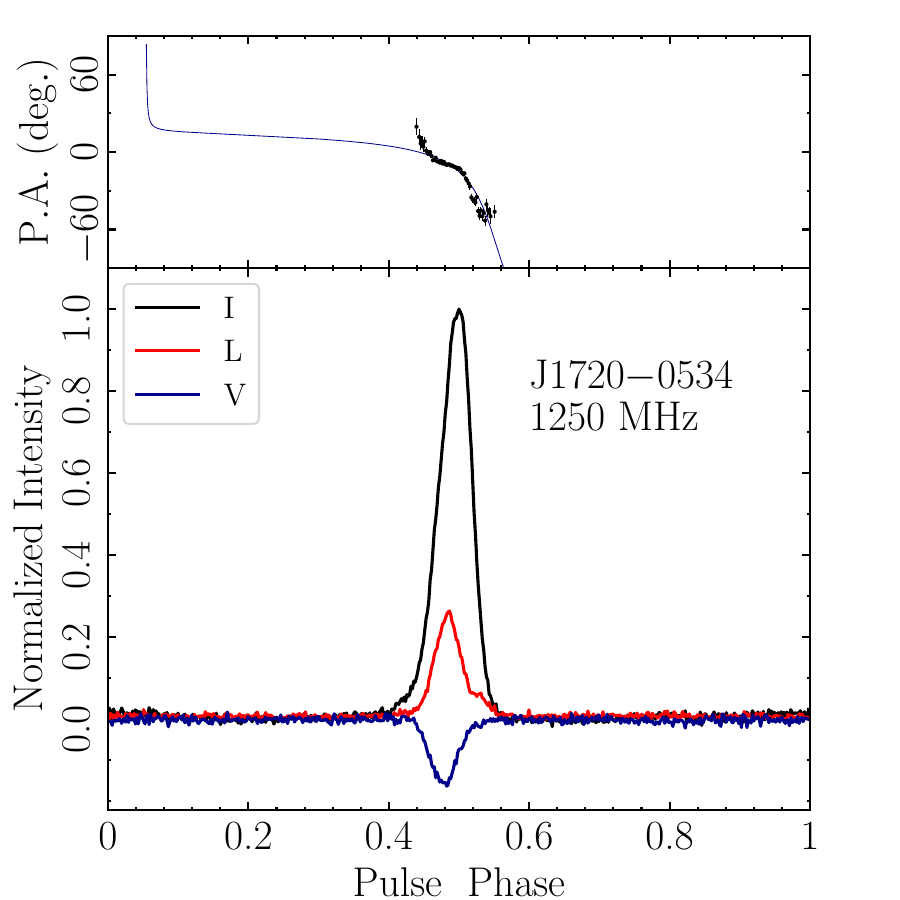}
\caption{
Pulse profile and polarization position angle for PSR~J1720$-$0534 at 350~MHz (left) and 1250~MHz (right).
The upper panel of each plot is the position angle, and the lower panel of is the polarization profiles.
The blue solid line in the upper panel represents the best-fitting RVM model.
The total intensity ($I$), linear polarization ($L$), and circular polarization ($V$) profiles are represented by black, red, and blue solid lines.
}
\label{fig:profile}
\end{figure}

Fig.~\ref{fig:res} shows the post-fit timing residuals as the function of orbital phases ($\phi_{b}$) and MJDs, with different colors representing different telescopes.
The eclipse and its surroundings have been excluded ($0.16<\phi_{b}<0.35$, by eyes).
The post-fit weighted RMS (WRMS) of timing residuals is 3.093~$\mu$s after applying an error factor (EFAC) of 1.43.

We present the variation of the excess DM (${\Delta}$DM) and the inferred excess electron column density ($N_{e}$) of 10 eclipse observations in Fig.~\ref{fig:dmx}, with different colors representing different observations.
These observations, which cover the phase of the eclipse, were all made with the FAST telescope.
On MJD~58987 and MJD~59262, we detect two significant DM fluctuations during the eclipse egress (called blips, \citealt{Crowter+2020MNRAS}).
The close-ups of the blips are presented in the insert plot of Fig.~\ref{fig:dmx}.
The pulse profile and linear polarization position angle (PA) for PSR~J1720$-$0534 at 350~MHz (MJD~59626) and 1250~MHz (MJD~59313) are shown in Fig.~\ref{fig:profile}.
The pulse profile at 820~MHz is similar to the 1250~MHz one.
As described in Rotating Vector Model (RVM, \citealt{Radhakrishnan1969, Everett2001}), the linear polarization PA is a function of the magnetic inclination angle ($\alpha$), the angle between the line of sight and the rotation axis ($\zeta$), the reference position angle ($\psi_{0}$), and longitude of the fiducial plane ($\phi_{0}$).
We used the RVM model to model the viewing geometry of this pulsar.
The geometric parameters used for RVM fits are $\alpha=84.2^{\circ}$, $\zeta=96.1^{\circ}$, $\psi_{0}=-74.2^{\circ}$ and $\phi_{0}=221.8^{\circ}$.

\begin{figure}
\centering
\includegraphics[width=0.95 \columnwidth]{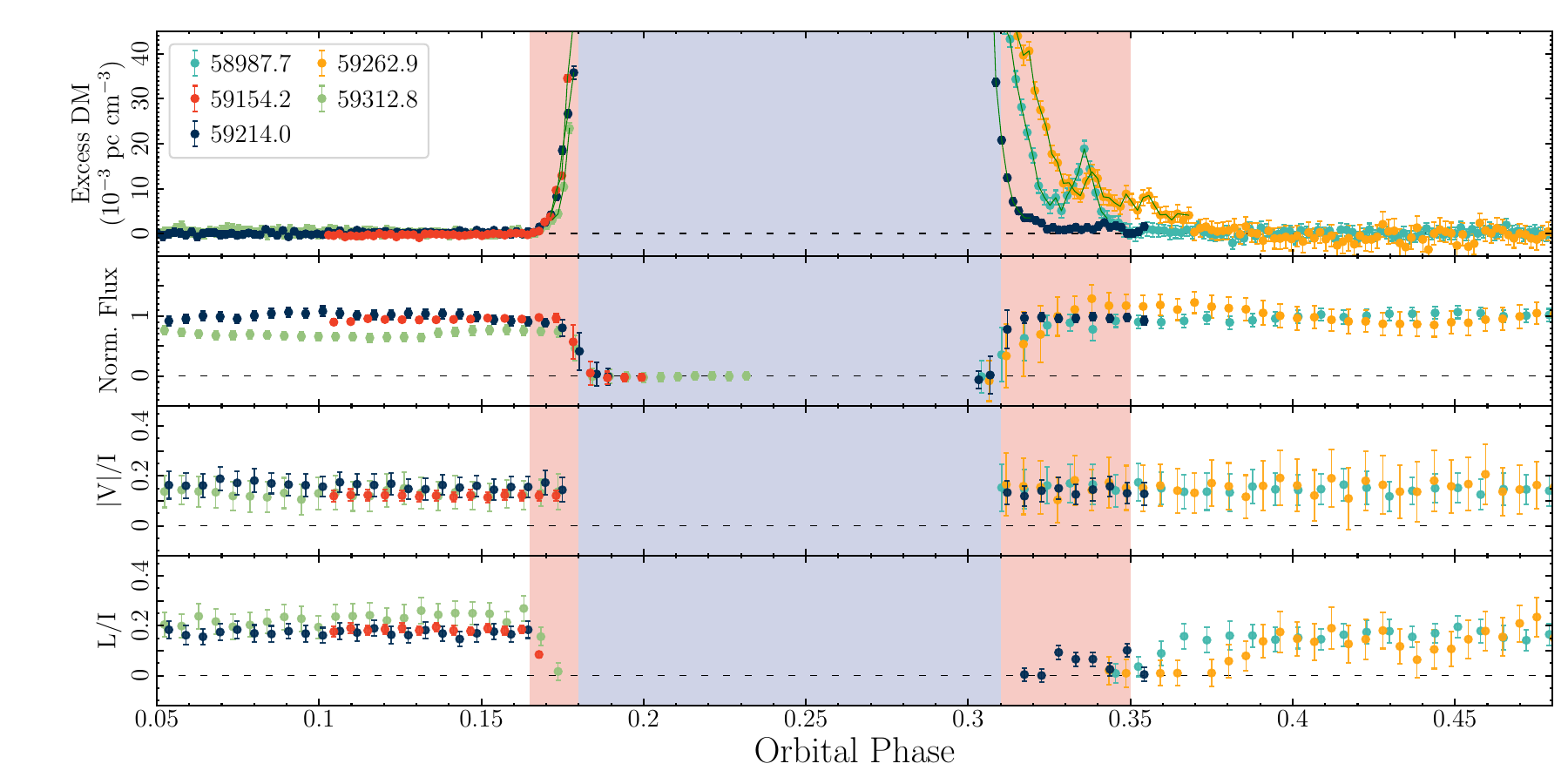}
\caption{
Variation of the ${\Delta}$DM, normalized mean flux density, and polarization fractions along the eclipse.
Different colors represent different observations.
The the upper panel with $\Delta \rm{DM}\ge 0.006$~$\rm{\text{pc}\,\text{cm}^{-3}}$ in same observation are connected by lines.
The eclipse region is shaded in purple at the orbital phase between 0.18 and 0.31.
The linear depolarization region is shaded in pink at  orbital phases  of 0.165--0.180 and 0.31--0.35.
}
\label{fig:depol_phase}
\end{figure}

The variations of ${\Delta}$DM, normalized flux density and polarization fraction for five observations covering the eclipses are shown in Fig.~\ref{fig:depol_phase}.
At $\phi_{b}$ between 0.165 and 0.180, the circular polarization is detected while the linear polarization is depolarized.
The circular polarization starts to be detected at $\phi_{b}=0.31$ while the linear polarization is still depolarized.
The circular polarization is eclipsed at the similar orbital phase as the total intensity, as a result the linear polarization presents a wider eclipse range, disappears earlier, and appears later.

\begin{figure}
\centering
\includegraphics[width=0.95 \columnwidth]{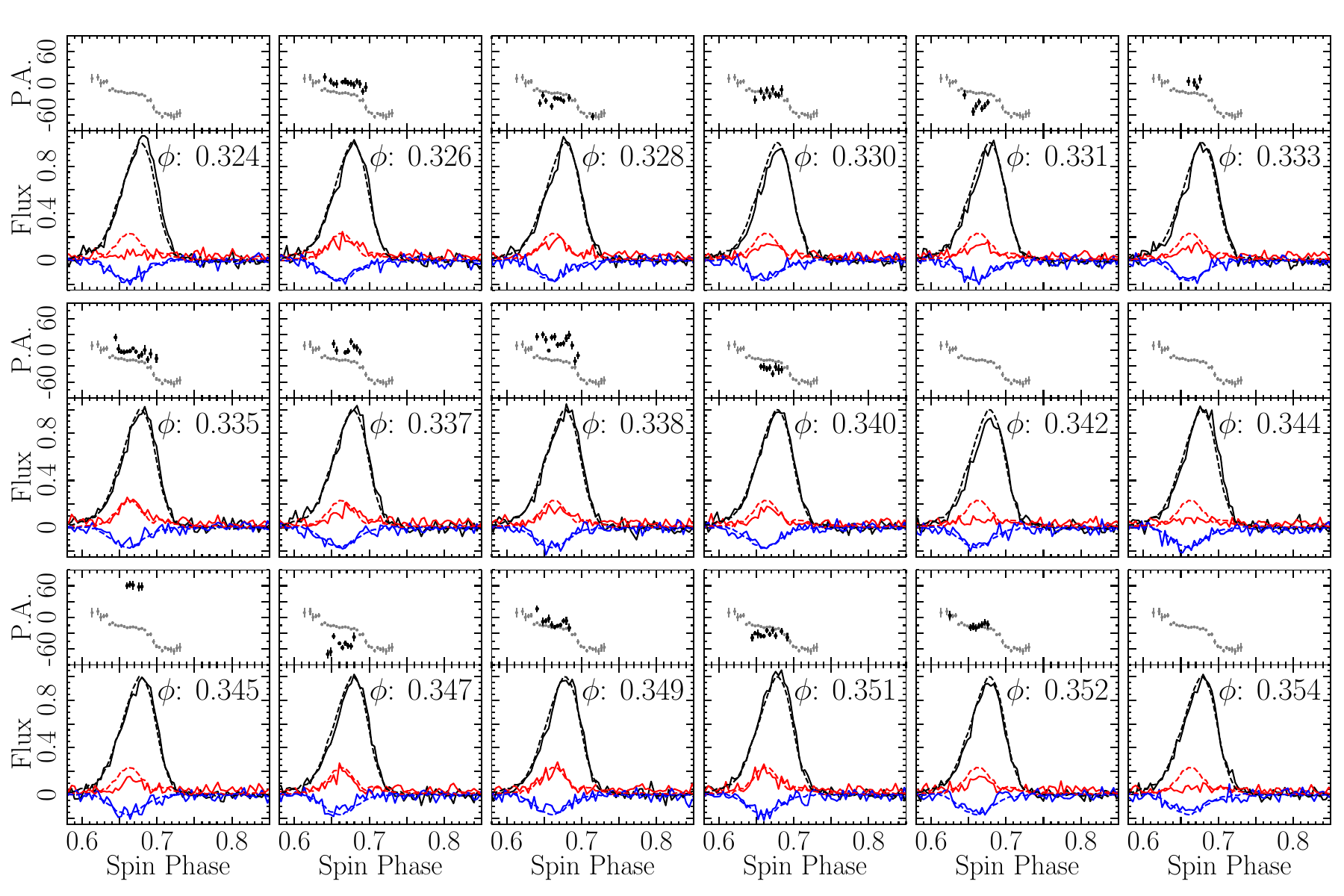}
\caption{Sub-integration polarization profile variation in different orbital phases.
The sub-integration profiles are for the integration time of 20 seconds for a FAST telescope observation on MJD~59214.
The orbital phase ($\phi$) at the center of each integration is at the top right corner of each subplot. 
The lower panel of the sub-plot gives the sub-integration pulse profile (solid line) comparing with the average pulse profile (dashed line) from this observation, with I, L, V profiles are plotted in black, red, and blue.
the different colors representing different polarization. 
The position angle (PA) of the average profile comparing with the sub-integration profile are presented at the top panel.}
\label{fig:depolprofile}
\end{figure}

\begin{figure}
\centering
\includegraphics[width=0.95 \columnwidth]{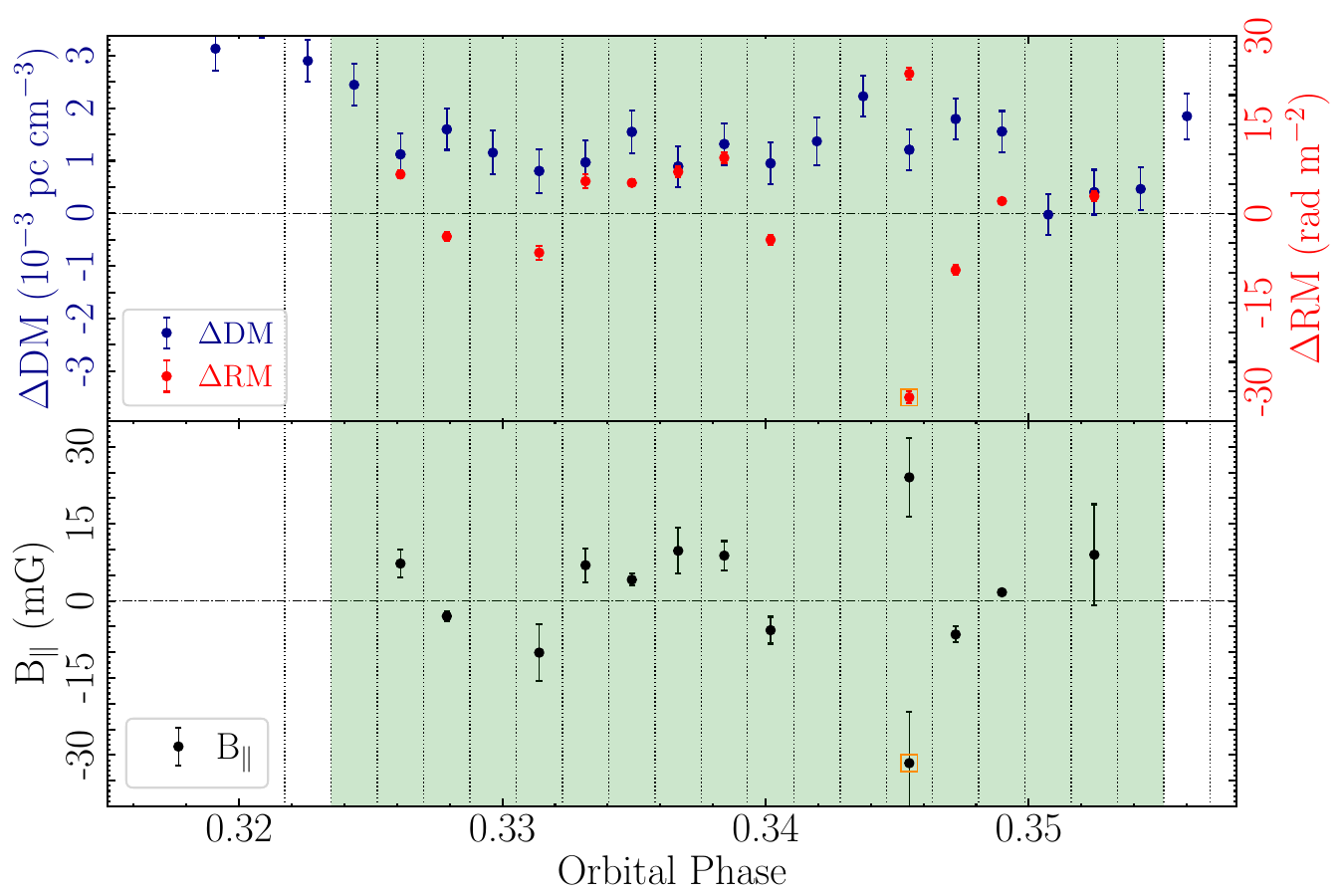}
\caption{
The detailed ${\Delta}$DM (blue dots), ${\Delta}$RM (red dots), and extra magnetic field strength at the LOS direction ($B_{\parallel}$, black dots).
Dotted vertical lines show the division of each sub-integration and dashed  horizontal lines at 0 are added for comparison.
At the orbital phase of 0.345, there are two possible values for ${\Delta}$RM.
The ${\Delta}$RM and extra magnetic field derived from the ${\Delta}$PA greater than $90^\circ$ are marked with orange squares.}
\label{fig:dmx_and_magenetic}
\end{figure}

Only at MJD~59214, we observed a shift in the PA when the linear polarization appears again in the orbital phase of 0.32 -- 0.35 (shown in Fig.~\ref{fig:depolprofile}).
The flux of linear polarization shows a decrease at random decrease at different orbital phase and totally disappeared at $\phi_{b} =$ 0.324, 0.330, 0.342, 0.344, and 0.354.
The PA profiles show the random shifts, as well.
At the same time, the total intensity profiles ($I$) and circular polarization profiles ($V$) show no deviations.
The $\Delta \rm{DM}$ and $\Delta \rm{RM}$ measurements between the orbital phase 0.32 and 0.35 are presented in the top panel of Fig.~\ref{fig:dmx_and_magenetic}.

Recent research indicates that a magnetic field permeates the eclipse material being the cause of a shift in PA and linear depolarization \citep{you+2018ApJ,Crowter+2020MNRAS,li+2022arXiv220507917L}.
The average magnetic field in the eclipse material $\langle B_{\parallel}\rangle$ is given by Equation \ref{equ:eq1} while assuming the $\rm{\Delta RM}$ and $\rm{\Delta DM}$ are originating in the same place \citep{Crowter+2020MNRAS, li+2022arXiv220507917L}.
Positive/negative RMs indicate the direction of {\Bmean} is towards/away from the observer.
\begin{gather}
\label{equ:eq1}
\langle B_{\parallel}\rangle = 
\rm{1.232~\mu G \left(\frac{\Delta RM}{rad~m^{-2}}\right)\left(\frac{\Delta DM}{pc~cm^{-3}}\right)^{-1}}
\end{gather}

Detailed {\Bmean} measurements are presented as black dots in the bottom panel of Fig.~\ref{fig:dmx_and_magenetic}.
These $\Delta \rm{DM}$ and $\Delta \rm{RM}$ imply a maximum magnitude of {\Bmean} towards the LOS direction of $\sim24\pm8$~mG and a maximum {\Bmean} $\sim-10\pm5$~mG in the opposite direction.

It is worth noting that when $|\Delta \rm{PA}| \ge 90^{\circ}$, it is possible that the PA would wrap around and appear on the opposite side of the baseline. This could lead to a misinterpretation of the $\Delta \rm{RM}$. 
However, at 1.25~GHz, a $\Delta \rm{RM}=27~\rm{rad~m^{-2}}$ that lead to $|\Delta \rm{PA}|$ of $90^{\circ}$ will also reduce the linear polarization fraction $L_{\rm obs}$ to 63\% of the original level ($L_{\rm ori}$) \citep{li+2022arXiv220507917L}.
But the polarization reduction fractions ${L_{\rm obs}/L_{\rm ori}}$ measured from most of the egress phases are greater then 70\%, which means that the $|\Delta \rm{PA}|$ at these phases should be smaller than $90^{\circ}$, except for $\phi_{b}=0.345$, where ${L_{\rm obs}/L_{\rm ori}} \sim 63\%$ (the left bottom panel of Fig.~\ref{fig:depolprofile}). 
At the orbital phase of 0.345, there are two possible values for $\Delta \rm{PA}$: $78\pm3^\circ$ or $-102\pm3^\circ$.
Theory predicts that ${L_{\rm obs}/L_{\rm ori}}$ should drop to 55\% if the $\Delta \rm{PA}$ is $-102^{\circ}$.
This is in tension with the observed value. 
Therefore, we think that $\Delta \rm{PA}=78\pm3^\circ$ is favoured for $\phi_{b}=0.345$.

\section{Discussion \& Conclusion}\label{sec:discuss}
In the two-year timing campaigns, we measured the astrometric parameters (RA and DEC), pulsar parameters (${\nu}$, ${\dot{\nu}}$, and DM), and binary parameters ($P_{\rm{B}}$, $x$, $T_{asc}$, $\epsilon_{1}$, and $\epsilon_{2}$) of PSR~J1720$-$0534 (Tab. \ref{tab:par}).
The measurements of the proper motion, projected semi-major axis derivative, and orbital period derivative are future objectives that could help us reveal more about the binary evolution of this system.

PSR~J1720$-$0534 is a BW pulsar system and is eclipsed at the orbital phases of $0.19-0.31$ ($\sim$12$\%$ of its orbit) at L-band.
Assuming $i=90^{\circ}$ and a pulsar mass of 1.35~$M_{\odot}$, the minimum pulsar-companion separation is $\sim$1.2~$R_{\odot}$ and the radius of the companion's Roche lobe \citep{Eggleton+1983ApJ} is $R_{L}^{c}=0.16~R_{\odot}$.
Combining the eclipse span and pulsar-companion separation, the eclipse radius $R_{E}$ is 0.45~R$_{\odot}$~($\sim$2.8~$R_{L}^{c}$) \citep{Altamirano+2011ApJ}.

The intensity eclipses of PSR~J1720$-$0534 are consistent with a symmetrical core plus variable edges.
The ingresses happen at $\phi_{b}=0.18$ and last for $\sim$1\% of the orbit (half-maximum flux density).
The egresses happen at the $\phi_{b}=0.31$ and the egress duration varies between 1$-$2\% of the orbit.
The steady ingress and swept-back egress suggest that the eclipse material around the companion leaves a comet-like tail due to orbital motion.

The strength of these fields is similar to the turbulent and ordered field observed from other BW pulsar systems inferred using Faraday rotations \citep{you+2018ApJ,Polzin+2019MNRAS,Crowter+2020MNRAS,li+2022arXiv220507917L}.
The oscillatory magnetic field in egress is observed for the first time and presents an important piece of evidence for understanding the interactions between the pulsar and the companion. 

The polarization observed in the egress phase reveals an orderly magnetic field showing an oscillatory pattern with a time scale of 20 seconds and direction reversal.
This oscillation time scale corresponds to a length scale of $\sim5\times10^3$~km when assuming an $i$ of 90 degrees.

We summarize the eclipsing event as followed:
\begin{itemize}
  \item A symmetrical complete-eclipsing body larger than the companion star in physical dimension;
  \item Varying asymmetrical ingress and egress, through which the pulsar signal can pass with different levels of extra dispersion and depolarization.
  \item Electron densities between 0 and $> 1.3 \times 10^7$~cm$^{-3}$ were found in the ingress and egress.
  \item We find a 10-milliGauss-level reciprocating magnetic field from one of the eclipse egresses with a low electron density ($<10^6$~cm$^{-3}$). 
\end{itemize}

Various physical pictures have been proposed in the literature to explain the eclipse of black widows and redbacks. 
In this paper, we explain our observations of J1720$-$0534 with a particular picture that was proposed as one of the possible scenarios for other BW pulsars (e.g \citealt{Phinney1988Natur,Thompson+1994ApJ, Wadiasingh+18}).
We suggest that the eclipse is caused by the magnetosphere of the brown dwarf.
The brown dwarf magnetosphere interacts with the pulsar wind in a similar way as the Earth magnetosphere interacts with the Solar winds \citep{Sckopke+83}.
There is a shock boundary between the companion magnetosphere and the pulsar wind outside.
The 10-milliGauss-level reciprocating magnetic field observed in the egress is part of the pulsar wind outside of the magnetosphere. 

\begin{figure}
\centering
\includegraphics[width=0.95 \columnwidth]{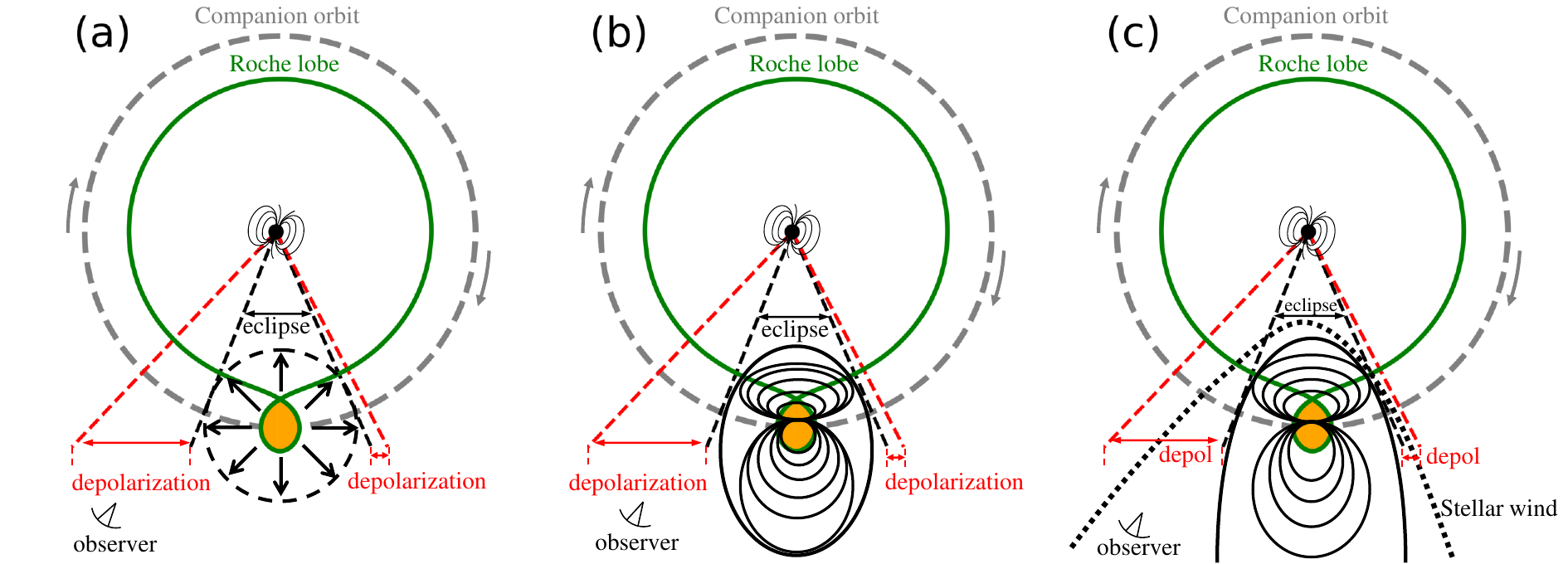}
\caption{Three general categories of models for the eclipsing body: (a) a naive companion wind picture; (b) the companion magnetosphere; (c) the companion's magnetosphere plus a pulsar wind shock boundary. The Roche lobe and companion orbit are calculated assuming a pulsar mass of 1.35~$M_{\odot}$ and an orbital separation of 1.2~$R_{\odot}$ ($i=90^{\circ}$). The Roche lobe of the companion star is filled in yellow. The eclipse edges and linear depolarization edges are shown by black and red dashed lines. The black dashed curves represent the magnetic field and the black solid curves represent the edge of stellar wind. The grey arrows indicate the companion's direction of motion.}
\label{fig:illu}
\end{figure}

To better illustrate our idea, we present the following three general categories of models, discuss their implications, and provide evidence that falsifies some of them and supports the other (Fig.~\ref{fig:illu}).

\subsection*{Naive stellar wind}

\begin{figure}
\centering
\includegraphics[width=0.95 \columnwidth]{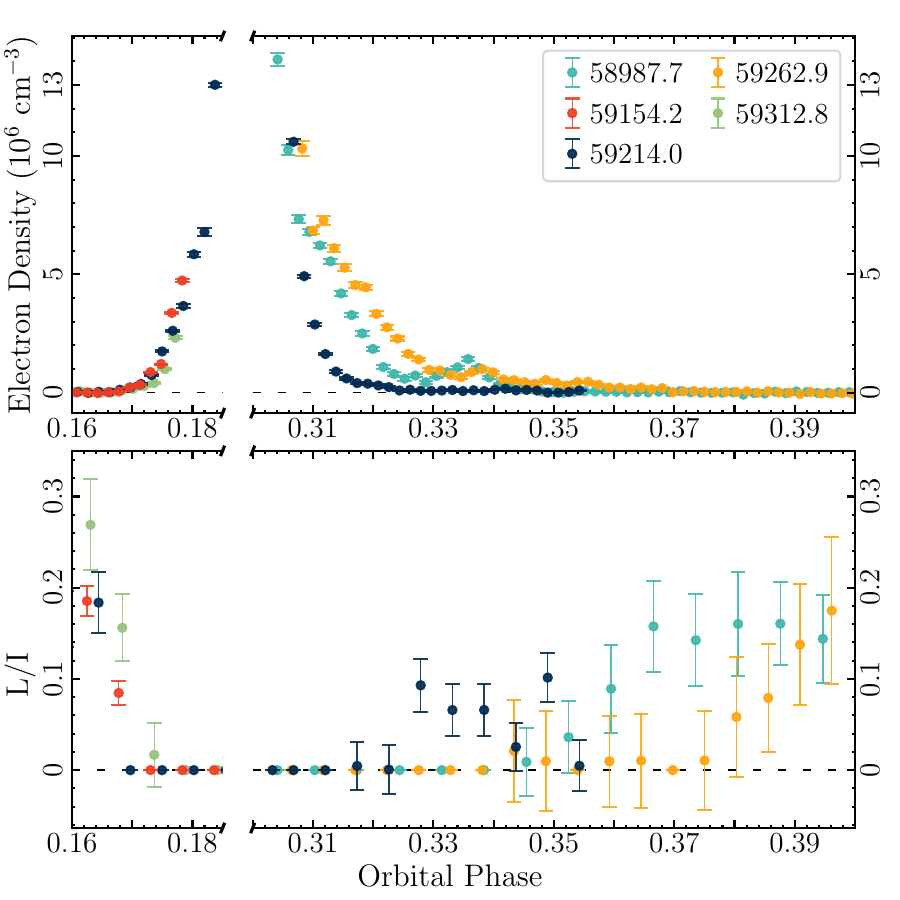}
\caption{
Upper panel: electron volume density at ingresses and egresses as a function of eclipse phase $\phi_{b}$: $n_e (\phi_{b}) = 0.5n_{H}R_{E}^{-1}\sin{\phi_{b}}^{-1} (\sin{(\pi/2-\phi_{b})}-\frac{1}{3}\sin{(\pi/2-\phi_{b})}^{3})^{-1}$ assuming a wind density profile of $n_e\propto 1/r$, where $r$ is the distance from the companion center, $R_{E}$ is the eclipsing radius and $n_H$ is the column density derived from the excess DM (Fig.~\ref{fig:dmx_and_magenetic}).
Lower panel: A close-up of the liner polarization fraction at the eclipse ingresses and egresses.
}
\label{fig:electron_density}
\end{figure}

An important feature of the black widow eclipse is the fact that the eclipsing body is bigger than the companion star's Roche lobe ($R_{E}\sim2.8~R_{L}^{c}$).
When \cite{Fruchter+1988} discovered the first black widow system B1957+20, they suggest that the eclipse could be caused by a stellar wind from the companion evaporated by the pulsar.
Fig.~\ref{fig:illu} illustrate an eclipsing body with an isotropic out-flowing stellar wind that surrounds the companion.
The speed of the stellar wind should be far greater than the orbital velocity of the companion $V_{\rm orb}=409$~km/s ($i=90^{\circ}$) in order to create a symmetrical eclipsing body.
However, a stable isotropic wind should have a density profile of $\rho(r) \propto 1/r$ based on Gauss's theorem.  
Such a density profile is smooth towards the eclipsing edge, inconsistent with what was observed at the edge of the eclipsing body (Fig.~\ref{fig:electron_density}).

Note that electron density $n_e$ is often estimated simply by taking the column density $n_H$ derived from dispersion and dividing  it with the eclipsing body diameter $2R_{E}$. This is a crude estimation that does not take the distance from the eclipse center into account. 
In Fig.~\ref{fig:electron_density} we present $n_e$ based on $\propto 1/r$ profile but account for the different LOS length for different eclipsing phase $\phi$. 
This figure helps us demonstrate the discrepancy between a wind density profile and the observed density profile. 
But the presented $n_e$ should also serve as a reasonably good general estimation for electron density for our later discussion.

\subsection*{The companion magnetosphere}

In the other scenario (Fig.~\ref{fig:illu}b), the pulsar emission is eclipsed by the plasma in the magnetosphere of the companion.
Assuming the companion brown dwarf is Roche Lobe-filling and has a surface magnetic field of $10^3$~G \citep{Reiners+2010AA}, we get a magnetic field strength of 4.5$\times 10^{1}$~G at the eclipse edge ($\phi_{b}=0.31$).
A magnetosphere field strength of 10~G is sufficient to trap and dominate plasma (assuming protons and electrons) of number density $<3\times10^{13}$~cm$^{-3}$ and velocity $\sim V_{\rm orb}$.

\cite{Thompson+1994ApJ} suggests that, at the edge of the magnetosphere of the brown dwarf, the magnetic pressure should balance the pulsar wind pressure, while the pulsar wind energy density is $U_{E}=\frac{\dot{E}}{4{\pi}ca^{2}}$ and the magnetic pressure is $\frac{B_{E}^{2}}{8{\pi}}$. The $\dot{E}$ is the spin-down luminosity, c is the speed of light, $a$ is the orbital separation, and $B_{E}$ is the magnetic field of the eclipse medium.
From this, the magnetic field strength of $B_{E}$ should be $\approx8~$G \citep{wang+2021ApJ}.

Interestingly, the derived theoretical magnetic field strength (45~G) is more than sufficient for the required field strength (8~G) at the eclipsing edge. 
However, these field strengths are more than three orders of magnitude higher than the value observed in our egress (10~mG).


\subsection*{Pulsar wind}

The third scenario (Fig.~\ref{fig:illu} c) supplements the second one with pulsar wind and a shock boundary, and fixes the inconsistency mentioned above.
Such a picture was proposed by \cite{Phinney1988Natur} as one of the early models.
In this picture, a shock boundary exists between the magnetosphere and the pulsar wind. 
Outside of the shock boundary are high-speed, low-density pulsar wind particles traveling with a low magnetic field, and inside, the slow-moving, high-density plasma trapped by the companion's magnetic fields.  
This is similar to the boundary shock observed from the Solar wind and the Earth magnetosphere \citep{Sckopke+83} where both the electron density and magnetic field rose suddenly as the ISEE-1\footnote{International Sun-Earth Explorer 1} probe traveled downstream of the Solar wind into the Earth magnetosphere.

The majority of energy in the pulsar wind is carried by relativistic particles.
The magnetic fields in the pulsar wind could be much smaller than the magnetic field of the companion at the orbital distance. After all, the pulsar's magnetic field is only 1.6$\times10^8$~G at its 10~km radius surface (Tab.~\ref{tab:par}).
The pulsar wind is almost transparent to the pulsar emission. This is because of the low density and the high Lorenz factor of the wind particles.
The wind particles have motion masses far exceeding their rest masses, causing their Faraday rotation effect to be negligible (\citealt{qg00, whl11}). 
When a moderate amount of slow-moving ionized materials from the companion's magnetosphere flow out of the boundary and come to the pulsar wind side, the combination of the extra slow electrons and a reasonably low magnetic field (10~mG) environment leads to the incomplete depolarization and the Faraday rotation.
As we mentioned in the previous section, such a condition is rarely met (only be observed in MJD 59214).
In most of the ingresses and egresses of this pulsar, the out-flowing electrons are either too dense or too variable and often completely depolarize the pulsar signal.

\cite{Thompson+1994ApJ} predicted that the pulsar wind could contain an oscillating part around the eclipsing edge with an oscillation length of $cP/2\simeq500$~km, where $c$ is the speed of light and $P$ is the spin period of the pulsar.
It should be noted that such reciprocating magnetic fields in the pulsar wind was already illustrated in the model of \cite{Phinney1988Natur}.
But such field was never observed until now.
We observed an oscillation length of 5$\times10^3$~km, different from the prediction, possibly due to our viewing angle or the reconnection loop expanding due to pressure changes.

The pulsar wind is expected to contain an alternating magnetic field in both the radial and transverse directions. 
This means that the eclipse was due to plasma occupying a relatively small volume of space in which the field is relatively uniform in the radial direction.
It is possible that in the particular egress of MJD 59214, the eclipsing plasma was a small stream of matter escaping the magnetosphere.
This might not be the case for the other egresses, through which significantly higher electron densities were observed and pulsar signals are completely depolarized.

The asymmetry between the depolarizing ingress and egress phase could be caused by an asymmetrical boundary shock layer deformed by orbit motion as illustrated in Fig.~\ref{fig:illu} c.
In our picture, the shock between the magnetized plasma is collisionless and could have boundaries with varying thickness depending on the conditions at the eclipse egresses.

We conclude that the companion magnetosphere plus pulsar wind picture could explain most of the observations. 
It also predicts that a shock boundary exists between the pulsar and the companion.
Inside the shock boundary, the magnetic field could reach the $\gtrsim$10~G level required for pressure balance and intensity eclipse, outside the boundary both the electron density is $\sim 10^6$~cm$^{-3}$ and the magnetic field is only $\sim10$~mG. 
As suggested by \cite{Wadiasingh+18}, if sufficient dissipation and heating exist in this boundary, one may be able to observe double-peaked X-ray modulation like those observed in some low-mass millisecond pulsar binaries.

\section*{ACKNOWLEDGEMENTS}
We would like to express our appreciation to the anonymous reviewer.
This work has used the data from the Five-hundred-meter Aperture Spherical radio Telescope (FAST), Parkes 64m "Murriyang" radio-telescope, and Green Bank Telescope (GBT).
FAST is a Chinese national mega-science facility, operated by the National Astronomical Observatories of Chinese Academy of Sciences (NAOC).
The Parkes Radio Telescope (Murriyang) is part of the Australia Telescope National Facility, which is funded by the Australian Government for operation as a National Facility managed by CSIRO.
The Green Bank Observatory is a facility of the National Science Foundation (NSF) operated under cooperative agreement by Associated Universities, Inc.
The National Radio Astronomy Observatory is a facility of the NSF operated under cooperative agreement by Associated Universities, Inc. 
This work is supported by
the National Nature Science Foundation (NSFC) grant No. 
12041303, 
12041304, 
11873067, 
12133004, 
12203045, 
12203070, 
12203072, 
12103013, 
U2031117, 
T2241020, 
the CAS-MPG LEGACY project and the National SKA Program of China No 2020SKA0120200, the Foundation of Science and Technology of Guizhou Province Nos. ((2021)023), the Foundation of Guizhou Provincial Education Department (No.KY(2021)303), the National Key Research and Development Program of China No. 2022YFC2205202, 2022YFC2205203, the Major Science and Technology Program of Xinjiang Uygur Autonomous Region No. 2022A03013-1, 2022A03013-3, 2022A03013-4, the National Key Research and Development Program of China No. 2022YFC2205203 and the 2021 project Xinjiang Uygur autonomous region of China for Tianshan elites and the Youth Innovation Promotion Association of CAS under No. 2023069.
P.~Wang acknowledges support from the Youth Innovation Promotion Association CAS (id. 2021055), CAS Project for Young Scientists in Basic Reasearch (grant YSBR-006) and the Cultivation Project for FAST Scientific Payoff and Research Achievement of CAMS-CAS.
J.-b.~Wang acknowledges support from Zhejiang Provincial Natural Science Foundation of China under grant Mo. LY23A030001. M. McLaughlin and V. Blackmon are supported by the NSF Physics Frontiers Center award number 2020265.
\input{ephem}

\bibliography{19C13}{}
\bibliographystyle{aasjournal}

\appendix
\section{Observation parameters and eclipsing events}\label{chap:app}
A table with the timing parameters of the observations and the eclipsing events. The "Observation MJD" column indicates the start time of the observation (to the minute), and the "Eclipsing" column indicates whether the eclipse was captured in this observation.

\setlength{\tabcolsep}{2pt}\begin{center}
    \begin{longtable}{cccccccc}
  \caption{Timing observations of J1720$-$0534 and eclipsing events.}
  \label{tab:obsLog} \\
        \hline 
      Telescope  & $f_{\rm{center}}$  &  $\rm{Bandwidth}$  & $n_{\rm{chan}}$ & $T_{\rm{samp}}$ & Observation MJD & Observation Length & Eclipsing\\
                 & (MHz)              &  (MHz)               &                 & (${\mu s}$) &   &  (minutes) & Y/N\\
        \hline 
			\endfirsthead
			\multicolumn{1}{l}{{\bfseries \tablename\ \thetable{} -- continued}} \\
		\hline 
      Telescope  & $f_{\rm{center}}$  &  $\rm{Bandwidth}$  & $n_{\rm{chan}}$ & $T_{\rm{samp}}$ & Observation MJD & Observation Length & Eclipsing\\
                 & (MHz)              &  (MHz)               &                 & (${\mu s}$) &   &  (minutes) & Y/N\\
        \hline 
			\endhead
		\hline 
            \multicolumn{3}{l}{{-- continued}} \\
			\endfoot
		\hline
			\endlastfoot
      FAST       & 1250           &  500                  & 1024            & 49.152   & 58987.7177  &  50  & Y \\
      FAST       & 1250           &  500                  & 1024            & 49.152   & 58990.6943  & 200  & Y \\
      FAST       & 1250           &  500                  & 1024            & 49.152   & 58993.6922  & 105  & Y \\
      FAST       & 1250           &  500                  & 1024            & 49.152   & 58998.6783  & 105  & N \\
      FAST       & 1250           &  500                  & 1024            & 49.152   & 59001.6781  &   5  & N \\
      FAST       & 1250           &  500                  & 1024            & 49.152   & 59061.5563  &   5  & N \\
      FAST       & 1250           &  500                  & 1024            & 49.152   & 59081.4489  &  10  & N \\
      FAST       & 1250           &  500                  & 1024            & 49.152   & 59094.4285  &  10  & N \\
      FAST       & 1250           &  500                  & 1024            & 49.152   & 59109.3841  &  10  & N \\
      FAST       & 1250           &  500                  & 1024            & 49.152   & 59121.3556  &  10  & N \\
      FAST       & 1250           &  500                  & 1024            & 49.152   & 59129.3276  &  10  & N \\
      FAST       & 1250           &  500                  & 1024            & 49.152   & 59140.2981  &  10  & N \\
      FAST       & 1250           &  500                  & 1024            & 49.152   & 59164.2306  &  10  & N \\
      FAST       & 1250           &  500                  & 1024            & 49.152   & 59186.2083  &  15  & N \\
      FAST       & 1250           &  500                  & 1024            & 49.152   & 59192.1847  &  15  & N \\
      FAST       & 1250           &  500                  & 1024            & 49.152   & 59205.1752  &   5  & N \\
      FAST       & 1250           &  500                  & 1024            & 49.152   & 59218.1414  &  15  & N \\
      FAST       & 1250           &  500                  & 1024            & 49.152   & 59231.0423  &  15  & N \\
      FAST       & 1250           &  500                  & 1024            & 49.152   & 59255.9701  &  10  & N \\
      FAST       & 1250           &  500                  & 1024            & 49.152   & 59267.9802  &  10  & Y \\
      FAST       & 1250           &  500                  & 1024            & 49.152   & 59304.8621  &  10  & N \\
      FAST       & 1250           &  500                  & 1024            & 49.152   & 59335.7548  &  15  & N \\
      FAST       & 1250           &  500                  & 1024            & 49.152   & 59427.5515  &   5  & N \\
      FAST       & 1250           &  500                  & 1024            & 49.152   & 59439.5063  &   5  & N \\
      FAST       & 1250           &  500                  & 1024            & 49.152   & 59443.4924  &   5  & N \\
      FAST       & 1250           &  500                  & 1024            & 49.152   & 59456.4333  &   5  & N \\
      FAST       & 1250           &  500                  & 1024            & 49.152   & 59466.3917  &   5  & N \\
      FAST       & 1250           &  500                  & 1024            & 49.152   & 59469.3994  &   5  & N \\
      FAST       & 1250           &  500                  & 1024            & 49.152   & 59476.4153  &   5  & N \\
      FAST       & 1250           &  500                  & 1024            & 49.152   & 59494.3424  &   5  & N \\
      FAST       & 1250           &  500                  & 1024            & 49.152   & 59509.3376  &   5  & N \\
      FAST       & 1250           &  500                  & 8192            & 98.304   & 59139.3354  &  15  & N \\
      FAST       & 1250           &  500                  & 8192            & 98.304   & 59154.2994  &  20  & Y \\
      FAST       & 1250           &  500                  & 8192            & 98.304   & 59158.2786  &  20  & Y \\
      FAST       & 1250           &  500                  & 8192            & 98.304   & 59212.1536  &  60  & N \\
      FAST       & 1250           &  500                  & 8192            & 98.304   & 59214.0841  &  60  & Y \\
      FAST       & 1250           &  500                  & 8192            & 98.304   & 59258.9811  &  15  & N \\
      FAST       & 1250           &  500                  & 8192            & 98.304   & 59262.9751  &  40  & Y \\
      FAST       & 1250           &  500                  & 8192            & 98.304   & 59297.8722  & 120  & N \\
      FAST       & 1250           &  500                  & 8192            & 98.304   & 59312.8341  &  60  & Y \\
      FAST       & 1250           &  500                  & 8192            & 98.304   & 59356.7170  &  60  & Y \\
      FAST       & 1250           &  500                  & 8192            & 98.304   & 59583.0829  &  60  & N \\
      GBT        & 350            &  105                  & 128             & 20.48    & 59626.6761  &  46  & N \\
      GBT        & 820            &  240                  & 128             & 10.24    & 59309.3510  &   5  & N \\
      GBT        & 820            &  240                  & 128             & 10.24    & 59312.4686  &   5  & N \\
      GBT        & 820            &  240                  & 128             & 10.24    & 59316.4761  &   4  & N \\
      GBT        & 820            &  240                  & 128             & 10.24    & 59607.5164  &  35  & N \\
      Parkes     & 2368           &  3328                 & 3328            & 128      & 59588.0406  & 160  & N \\
    \end{longtable}
\end{center}

\end{document}

%% file: ephem.tex
\begin{table*}
\begin{center}
\caption{Best-fit \textsc{tempo} timing parameters for pulsar J1720-0534.}
\label{tab:par}
\begin{threeparttable}   
\begin{tabular}{lc}
\hline
\hline
Pulsar name  &J1720-0534   \\
\hline
\textit{Measured Parameters}&  \\
Right Ascension, ${\alpha}$ (J2000) \dotfill&  17:20:54.505914(2)\\
Declination, ${\delta}$ (J2000) \dotfill&  --05:34:23.82233(14)\\
Spin Frequency, ${\nu}$ (${\text{s}^{-1}}$) \dotfill&  306.030167430858(4)\\
Spin frequency derivative, ${\dot{\nu}}$ (${\text{s}^{-2}}$) \dotfill&  $-7.6539(19)\times10^{-16}$\\
Dispersion Measure, DM (${\text{cm}^{-3}\,\text{pc}}$) \dotfill&  36.82987(10)\\
Bianry model \dotfill&  ELL1\\
Orbital Period, ${P_{\rm b}}$ (day) \dotfill&  0.13169857686(2)\\
Projected semi-major axis, ${x}$ (lt-s) \dotfill&  0.05961281(10)\\
Epoch of the ascending node, Tasc (MJD) \dotfill&  58941.85271857(4)\\
${e\sin{\omega}, \epsilon_{1}( 10^{-5} )}$ \dotfill&  0.2(2)\\
${e\cos{\omega}, \epsilon_{2}( 10^{-5} )}$ \dotfill&  --4.3(2)\\
Mass function, ${f(10^{-3}) M_{\odot}}$ \dotfill&  0.01311331(6)\\
&  \\
\textit{Fixed Parameters}&  \\
Solar System Ephemeris$^a$ \dotfill&  DE438\\
Reference epoch for ${\alpha}$, ${\delta}$, and ${\nu}$ (MJD) \dotfill&  58990.725174\\
Data span (MJD)\dotfill& 58987 -- 59626\\
Number of TOAs\dotfill&  2697\\
Fit $\chi^{2}$/number of degrees of freedom\dotfill& 5524.72/2688\\
Post-fit RMS of residuals ($\mu$s)\dotfill& 3.093\\
EFAC\dotfill& 1.43\\
&  \\
\textit{Derived Parameters}&  \\
Minimum companion mass$^b$, $m_{\rm{c,min}} (\rm{M}_{\odot})$ \dotfill&  0.029\\
Median companion mass$^c$, $m_{\rm{c,med}} (\rm{M}_{\odot})$ \dotfill&  0.034\\
Inferred eccentricity, e ${(10^{-5})}$ \dotfill&  4.3(2)\\
Galactic longitude, ${l}$ (${^\circ}$) \dotfill&  17.0667302186(7)\\
Galactic latitude, ${b}$ (${^\circ}$) \dotfill&  17.25223302(4)\\
Spin-down luminosity, ${\dot{E}}$ (${10^{33}\,\text{erg}\,\text{s}^{-1}}$) \dotfill&  9.2\\
Surface magnetic field, ${B_\text{surf}}$ (${10^{8}\,\text{G}}$) \dotfill&  1.6\\
Characteristic age, ${\tau_\text{c}}$ (Gyr) \dotfill&  6.3 \\
\hline
\end{tabular}
	\begin{tablenotes}
        \item[a] The timing model use the DE438 solar system ephemeris and is referenced to the TT(BIPM) time standard. Values in parentheses are the 1$\sigma$ uncertainty in the last digit as reported by \texttt{TEMPO}.
        \item[b] $m_{\rm{c,min}}$ is calculated for an orbital inclination of $i=90^\circ$ and an assumed pulsar mass of 1.35~$\rm{M}_{\odot}$ \citep{Hobbs+2006MNRAS}.
        \item[c] $m_{\rm{c,med}}$ is calculated for an orbital inclination of $i=60^\circ$ and an assumed pulsar mass of 1.35~$\rm{M}_{\odot}$ \citep{Hobbs+2006MNRAS}.
      \end{tablenotes}
    \end{threeparttable}
\end{center}
\end{table*} 